\documentclass[12pt]{article}

\usepackage{graphicx}
\pdfoutput=1

\title{ On the effects at colliding $\mu$ meson beams }

\author{ F.F.~Tikhonin }

\date{}

\begin{document}

\maketitle

\begin{center}
{ \it Joint Institute for Nuclear Research, Dubna, 141980, Russia }
\end{center}

\begin{abstract}
  Possible influence of the weak interaction on the
$\mu^+ + \mu^- \to \mu^+ + \mu^- $ scattering and
the $\mu^+ + \mu^- \to e^+ + e^- $ reaction, both
through the neutral lepton currents and the charged
ones (in the second order on weak constant), are
considered.
   The calculations show that $P$ - odd effects in the
mentioned processes would prove the existence of the
neutral currents which, in their turn, give the
principal basis for explanation of the mass difference
of a muon and an electron from the point of view of
\cite{1}.
\end{abstract}

1.~~At present in the elementary particle physics the region of
lengths of $10^{-15} cm$ is investigated.
On the way to the yet smaller distances, which undoubtedly will be an
object of experimental investigations, there is a scale of the so called
weak interactions \begin{eqnarray}
l_W = \sqrt{\frac{G_W}{hc}} \approx 0.61\cdot10^{-16} cm.
\end{eqnarray}

Physics effects at this distance represent a significant interest
by number of reasons. Colliding beams of designed
accelerators in not-so-far future will give possibility of approaching
to a great extent to the scale of weak interaction.

Below some effects of muon physics are considered  at colliding $\mu$ -
meson beams at high energies. In particular, such experiments will
give the possibility to check out the existence of neutral currents
in the weak interactions of the type $(\bar{\mu} \mu)$ and $(\bar{e} e)$.

In this respect it is interesting to evaluate the impact of higher
approximations on the weak constant. A possibility of experimental
approach to the elucidation of the higher approximation contributions
represents considerable principal interest for the field theory in general,
because the weak interactions are nonrenormalizable. The existence of
neutral currents (not necessary with the universal coupling constant)
should give the possibility, in principle, to explain mass difference
of electron and muon in the spirit of the paper~ \cite{1}, because
just such a type of interactions give contributions of different signs
to the masses of electron and muon, if their fields satisfy equations
$(i\hat p + m)\phi_I = 0$ and $(i\hat p - m)\phi_{II} = 0$, respectively.

2.~~Let's calculate the cross section, corresponding to diagrams of the Fig.1,
i.e. take in account the weak interactions of a muon
as well as the Electromagnetic one.
Denoting momenta of initial $\mu^-$ and $\mu^+$ particles
as $s_1$ and $t_1$ and of the final particles as $s_2$ and $t_2$,
respectively, write out  the required matrix element
\begin{eqnarray}
&&{\cal M} = -\frac{e^2}{q^2}\Bigl[\bar u(s_2) \gamma_{\mu}u(s_1)\Bigr]
\Bigl[\bar v(t_1)\gamma_{\mu} v(t_2)\Bigr] \\ \nonumber
&& + \frac{e^2}{k^2}\Bigl[\bar u(s_2)\gamma_{\mu}v(t_2)\Bigr]
\Bigl[\bar v(t_1)\gamma_{\mu} u(s_1)\Bigr] \\ \nonumber
&& + \frac{G}{\sqrt2}\Bigl[\bar u(s_2) \gamma_{\mu}(1+\gamma_5)v(t_2)\Bigr]
\Bigl[\bar v(t_1)\gamma_{\mu} (1+\gamma_5)u(s_1)\Bigr] ;
\end{eqnarray}
in the  c.m.s. we have in the relativistic limit $q^2 = -4E^2 \sin^2
\theta/2$ ,~~ $~~k^2 = 4 E^2$,
where $\theta$ is the scattering angle and $E$ is the initial energy
of colliding beams.
Effective cross-section is calculated by the formula
\begin{eqnarray}
d\sigma = \frac{(2\pi)^4 |{\cal M}|^2}{4 \Bigl[(s_1t_1)^2 -\mu^4 \Bigr]^{1/2}}
\frac{d \vec s_2}{(2\pi)^32s^0_2}\frac{d \vec t_2}{(2\pi)^3 2t^0_2}
\delta^{(4)}(s_2 +t_2 -s_1 - t_1),
\end{eqnarray}
and the differential cross section in the c.m.s. at large energies
 $\bigl[r_0 = \frac{\alpha}{\mu}=1.36\cdot10^{-15}cm\bigr]$ is:
\begin{eqnarray}
&&\frac{d\sigma}{d(\cos\theta)} = \frac{\pi r^2_0}{8}
\Bigl(\frac{\mu}{E}\Bigr)
^2\Bigl\{2\frac{1+\cos^4\theta/2}{\sin^4\theta/2}+ (1+\cos^2\theta)
- 4\frac{\cos^2\theta/2}{\sin^2\theta/2}  \\ \nonumber \\ \nonumber
&& + ~~{\cal E}^{(n)}(1 - \frac{1}{\sin^2\theta/2})
+{\cal E}^{(n)})(1 +\cos\theta)^2 \Bigr\}.
\end{eqnarray}
Here the parameter ${\cal E}^{(n)}$ is introduced,
which describes the extent
of influence of weak interactions on $\mu^-\mu^+$ scattering,
\begin{eqnarray}
{\cal E}^{(n)} = \frac{8GE^2}{e^2\sqrt{2}} = 6.2\cdot 10^{-4}\frac{E^2}{m_N^2},
\end{eqnarray}
where $m_N$ is the nucleon mass.

  Angular distributions of pure electromagnetic scattering (Bhabha one),
and distributions taking into account the "weak"
corrections are shown on the Fig.3; corresponding numerical values are
displaced in Table 1 for the muon energies in the c.m.s. of 25, 30,
and 50 GeV. The most notable deviations from the electromagnetic
distributions  are in the angle region of $70^{\circ}- 100^{\circ}$,
so the measurements, made with the enough precision, could reveal
an expected difference. In these calculations the usual constant
of weak interactions was used. Possibly, this is so
indeed (in pure leptonic processes), although the search of e.g.
the $K^{\circ} \to \mu^+\mu^-$ decay gives the following relation~\cite{2}:
\begin{eqnarray}
\frac{\Gamma(K^{\circ} \to \mu^+\mu^-}{K^{\circ} \to \mu \nu}
\leq 1.5\cdot 10^{-6} .
\end{eqnarray}

3.~~  A parity nononconservation effect in the process under investigation
will be exhibited most clearly in the appearance of the longitudinal
polarization of the scattered muons AT the unpolarized initial beams.
For the calculation of polarization degree we will not sum on the
spin directions of the scattered muon. The muon polarization 4-vector
at the considered energies can be written in the following form
\begin{eqnarray}
s \approx \frac{E_\mu}{\mu}(\vec n \cdot \vec {\xi})(1,\vec n),
\end{eqnarray}
where $\vec n = \frac{\vec s_2}{|\vec s_2|}$ and $\vec {\xi}$  is the
unit vector on the muon polarization direction in its rest frame. Again
it can be found that pure electromagnetic part of
cross-section don't change, and remaining terms acquire the common
factor $(1 - \cos \Theta)$, where $\Theta$ is the angle between
$\vec n$ and $\vec {\xi}$ .Now, for the longitudinal polarization
the following expression can be found:
\begin{eqnarray}
P_l = \frac{-{\cal E}^{(n)} ( 1 - \frac{1}{\sin^2\theta/2}
+{\cal E}^{(n)})(1 +\cos\theta)^2 }
 {2\frac{1+\cos^4\theta/2}{\sin^4\theta/2}+ (1+\cos^2\theta)
- 4\frac{\cos^2\theta/2}{\sin^2\theta/2}+{\cal E}^{(n)} ( 1 -
\frac{1}{\sin^2\theta/2}+ {\cal E}^{(n)})(1 +\cos\theta)^2 }
\end{eqnarray}
Polarization degrees for several scattering angle values are shown
in Table 1. Smallness of $P_l$ hampers its measurement,
but instability of muons simplifies situation, because the
muons polarization
can be revealed by considering the angular distributions of the decay
products. Note, that in the process $\mu^+ + \mu^- \to e^+ + e^-$
polarization can reach sizable values, but its measurement is rather
difficult. We will address this process later.

4.~~     In the case of finding the $P$-odd effects in the $\mu^+\mu^-$
scattering we must be sure, that their source is namely the neutral
currents. To this end let's evaluate the contribution of the weak
interactions at the second order through the charged currents, and
write out the matrix element, corresponding to diagrams of the Fig. 2
with the regularized distribution functions of neutrinos.
\begin{eqnarray}
&&{\cal M}^{(2)} = \frac{G^2}{4}\frac{1}{(2\pi)^8}\delta^{(4)}(s_1+t_1-s_2-t_2)
\frac{\mu^2}{E^2}\int\limits_0^{\Lambda^2}d\lambda_1
\int\limits_0^{\Lambda^2}d\lambda_2\int d^4k \\ \nonumber
&& \times \Biggl\{\Bigl[\bar u(s_2)\Gamma_{\nu}u(s_1)
\bar v(t_1)\Gamma_{\mu}v(t_2)\Bigr]
Tr\Bigl(\frac{\hat k}{(k^2-\lambda_1)^2}\Gamma_{\nu}
\frac{\hat{k} + \hat{s}_1 - \hat{s}_2}{\bigl[k^2+2k\cdot(s_1-s_2)
+(s_1-s_2)^2-\lambda_2\bigr]^2 } \Gamma_{\mu}\Bigr) \\ \nonumber
&& -\Bigl[\bar u(s_2)\Gamma_{\mu}v(t_2)\bar v(t_1)\Gamma_{\nu}u(s_1)\Bigr]
Tr\Bigl(\frac{\hat k}{(k^2-\lambda_1)^2}\Gamma_{\mu}
\frac{\hat{k} - \hat{s}_1 - \hat{t}_1}{\bigl[k^2-2k\cdot(s_1+t_1)
+(s_1+t_1)^2-\lambda_2\bigr]^2 } \Gamma_{\mu}\Bigr)\Biggr\}
\end{eqnarray}
The integrals in this expression have the following form
\begin{eqnarray}
I^{\alpha\beta}(s_1-s_2)=\int\limits_0^{\Lambda^2}d\lambda_1
\int\limits_0^{\Lambda^2}d\lambda_2\int d^4k \frac{k^{\alpha}}
{(k^2-\lambda_1)^2}
\frac{k^{\beta} + s_1^{\beta} - s_2^{\beta}}{\bigl[k^2+2k\cdot(s_1-s_2)
+(s_1-s_2)^2-\lambda_2\bigr]^2 }.
\end{eqnarray}
 By  expanding it as a power series and retaining only the main terms
in $\Lambda$ we obtain for both integrals in (9)
\begin{eqnarray}
I^{\alpha\beta}_2=I^{\alpha\beta}_1 = - \frac{i\pi^2}{4}\Lambda^2
g^{\alpha\beta}.
\end{eqnarray}

Scattering cross section corresponding to the diagrams of Fig. 2
has the form
\begin{eqnarray}
\frac{d\sigma}{d(\cos\theta)} = \frac{3(G\Lambda)^4 E^2}{2^3 \pi^5}
(1 + \cos\theta)^2 .
\end{eqnarray}
Defining from this the parameter ${\cal E}^{(ch)}$ corresponding to
the ${\cal E}^{(n)}$ introduced earlier, we can characterize the comparative
influence of charged and neutral weak currents on the muons scattering
by the parameter
\begin{eqnarray}
\xi_{W}(\Lambda) = \frac{{\cal E}^{(ch)}}{{\cal E}^{(n)}} =
12.4\cdot 10^{-7}(\frac{\Lambda}{m_N})^2 .
\end{eqnarray}
For the three values of $\Lambda$ (in units of nucleon mass) one obtains
\begin{eqnarray}
\xi_{W}(50) \cong 3.2 \cdot 10^{-3}, ~~~\xi_{W}(100) \cong 12.4 \cdot 10^{-3},
~~~\xi_{W}(300) \cong 10^{-1}.\end{eqnarray}

    There are some reasons to assume that the most probable value of
$\Lambda$ is less than 100. In this case values of $\xi_W$ indicate that
neutral currents are more preferable in the $\mu^+\mu^- \to \mu^+\mu^- $
 scattering if neutral Fermi constant is sufficiently large. However,
let us draw attention to the fact, that these conclusions obtained at
the presumption, that in the course of calculation in the second order
of perturbation theory on $G$ the main divergence was $\Lambda^2$,
although there is a possibility that in the consistent theory
this term  does not contribute ~\cite{3}. In this case arguments in the
favor of neutral currents are greatly enhanced. Thus, the most important
question in the nonrenormalizable theories on divergences expects
its direct experimental resolution. In view of great successes in the
developing of accelerator techniques at colliding beams this task is not
the subject of far away future.

5.~~Note, that the presence of exchanged diagram at the Fig. 1 has strong
influence on the scattering process causing the singular dependence on
the scattering  angle. From the other side this digram is absent in the
reaction $$\mu^+ + \mu^- \to e^+ + e^-$$ and one expects  more clear
picture of angular distribution of electrons at all values of $\theta$.
With the weak diagram taken into account one obtains the following
form of the cross section
\begin{eqnarray}
\frac {d\sigma}{d \cos\theta} =\frac {\pi}{8} \frac{\alpha}{E^2}
\Bigl[ 1 +\cos^2\theta +{\cal E}_1^{(n)}(1 +{\cal E}_1^{(n)}
(1+\cos\theta)^2)\Bigr].
\end{eqnarray}
Table 1 contains the numerical characteristics of this
process and Fig.4 shows the angular distribution of
electrons. A considerable influence of the weak
interaction on this process is noticeable.

   Now evaluate the second order on $G$ in the
reaction $\mu+\mu^- \to e^+e^-$; here
\begin{eqnarray}
\frac {d\sigma}{d\Omega} = \frac{4G^2 E^2}{(2\pi)^5}
(G\Lambda^2)^2(1 + \cos\theta)^2.
\end{eqnarray}
   The relative influence of the weak neutral and
charged currents is characterized by the same value of
$\xi_W$ as in the case of the process
$$\mu^++\mu^-\to \mu^++\mu^-$$.

6.~~  From the calculations made above it can be concluded,
that experiments with the colliding $\mu^+$ and $\mu^-$
beams would allow to reveal the influence of weak
interaction. Manifestation of a parity non-conservation
effects (with equal coupling constants of charged and
neutral currents) might be considered as revealing the
neutral currents of type $(\bar{\mu} \mu)$ and
$(\bar{e}e)$. And even in the case if $G^{neutral}
\simeq (10^{-2} - 10^{-3})G^{charged}$ this conclusion
will remains true.

 Let's stress, that the considerations made above are
based on the perturbation theory, which possibly is
inapplicable at the critical energy of weak interaction.
Approaching this critical region it is required
to take into account, in some form, a change of
the initial state (its damping).

Author sincerely  grateful to ac. M.A.Markov for
proposal to theoretically investigate effects at
the colliding beams and for permanent discussion
of all issues concerning this work.

\begin{verbatim}
                                      Received October 16, 1968.
\end{verbatim}

\newpage

\begin{center}
Table and figure captions.
\end{center}
\begin{itemize}
\item{Table 1.} Angular distribution and polarization
of final beam in the $\mu^+\mu^- \to \mu^+\mu^-$ scattering.
\item{Table 2.} Angular distribution and polarization
of final beams in the process $\mu^+\mu^- \to e^+ + e^-$.
\item{Fig 1.} Diagrams for the $\mu^+\mu^-$ scattering at tree
level are shown.
\item{Fig 2.} Diagrams for the $\mu^+\mu^-$ scattering at
one-loop are shown.
\item{Fig 3.} Angular distributions of the process $\mu^+\mu^-
\to\mu^+\mu^-$ are shown. Curve 1.: Bhabha angular distribution,
Curve 2.: distribution with the weak interactions taken into account.
\item{Fig 4.} Angular distributions of the process $\mu^+\mu^-
\to\mu^+\mu^-$ are shown. Curve 1. is for pure electromagnetic
process and  Curve 2. represents distribution with the weak interactions
taken into account.
\end{itemize}

\pagestyle{empty}

\newpage
\begin{figure}[htb]
\begin{center}
\resizebox{16cm}{!}{\includegraphics*{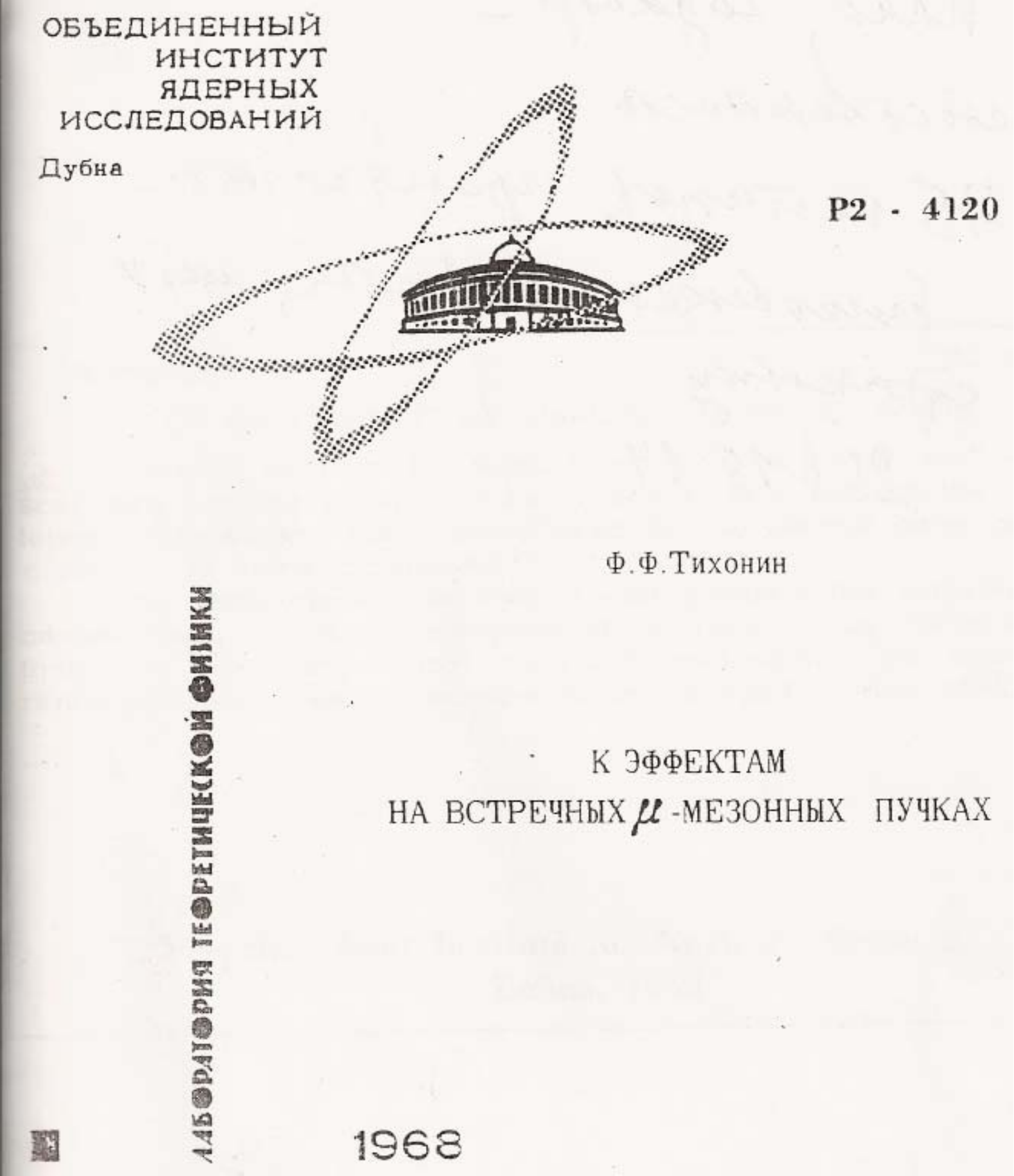}}
\end{center}
\end{figure}

\newpage
\begin{figure}[htb]
\begin{center}
\resizebox{16cm}{!}{\includegraphics*{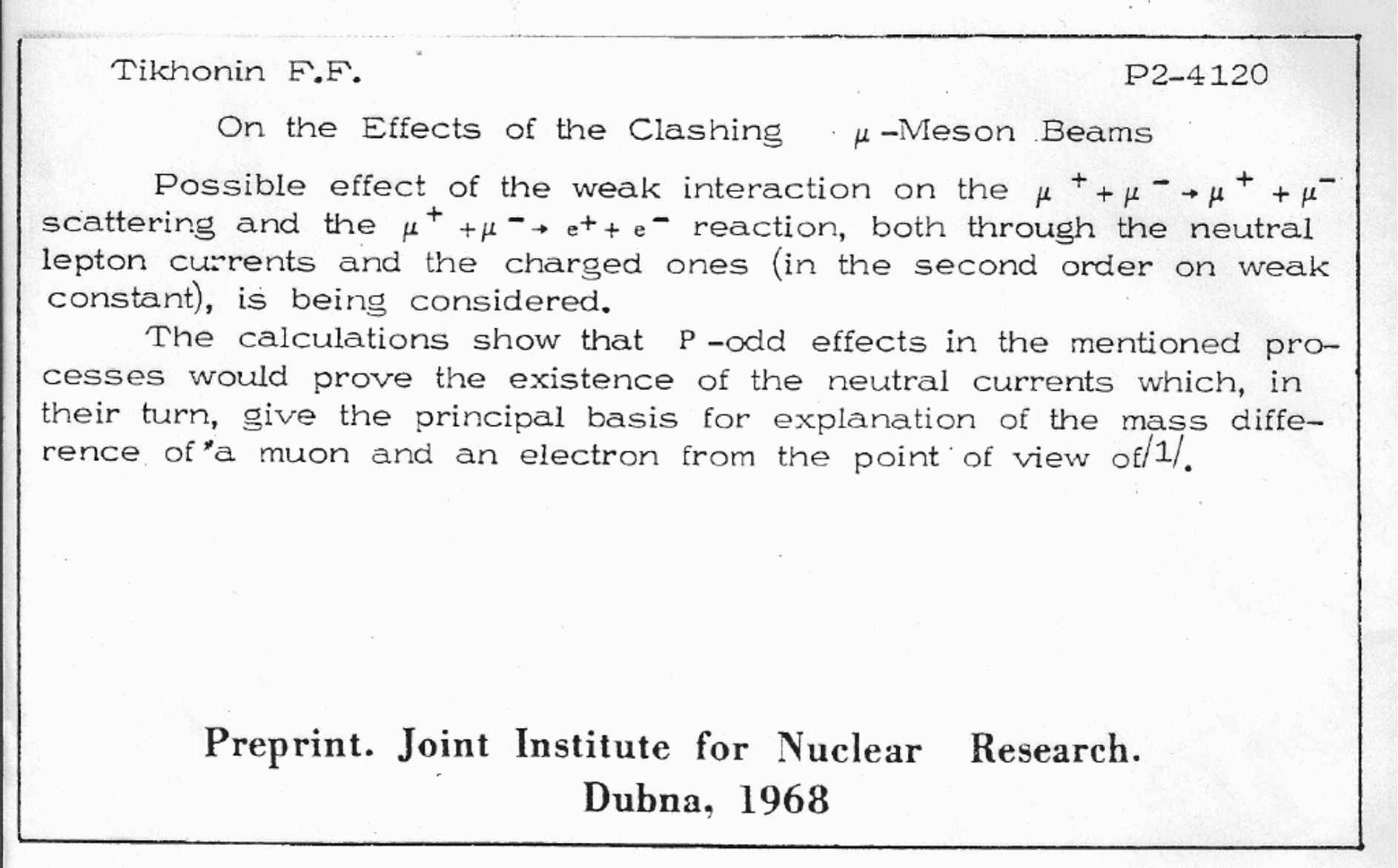}}
\end{center}
\end{figure}

\newpage
\begin{figure}[htb]
\begin{center}
\resizebox{16cm}{!}{\includegraphics*{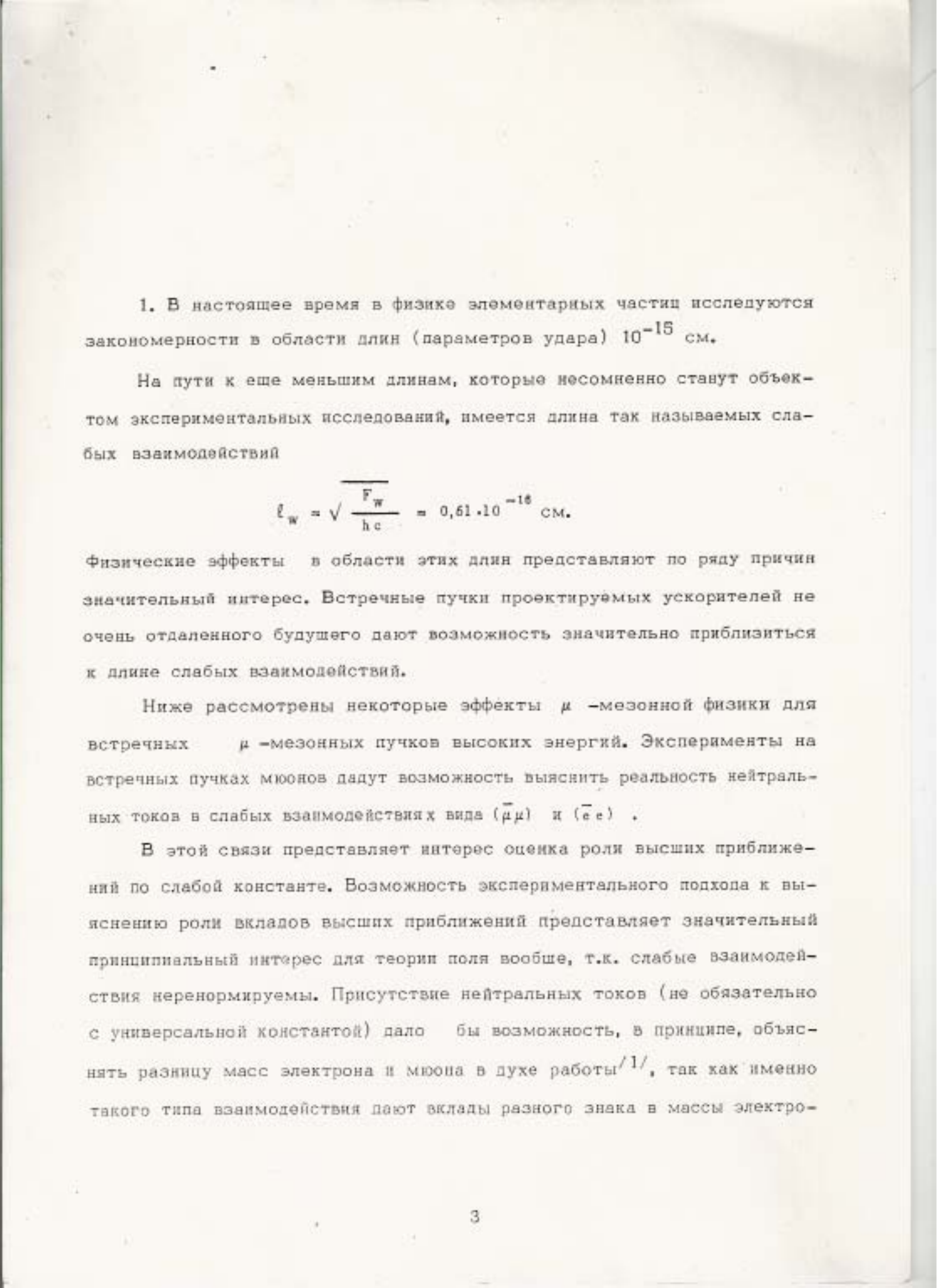}}
\end{center}
\end{figure}

\newpage
\begin{figure}[htb]
\begin{center}
\resizebox{16cm}{!}{\includegraphics*{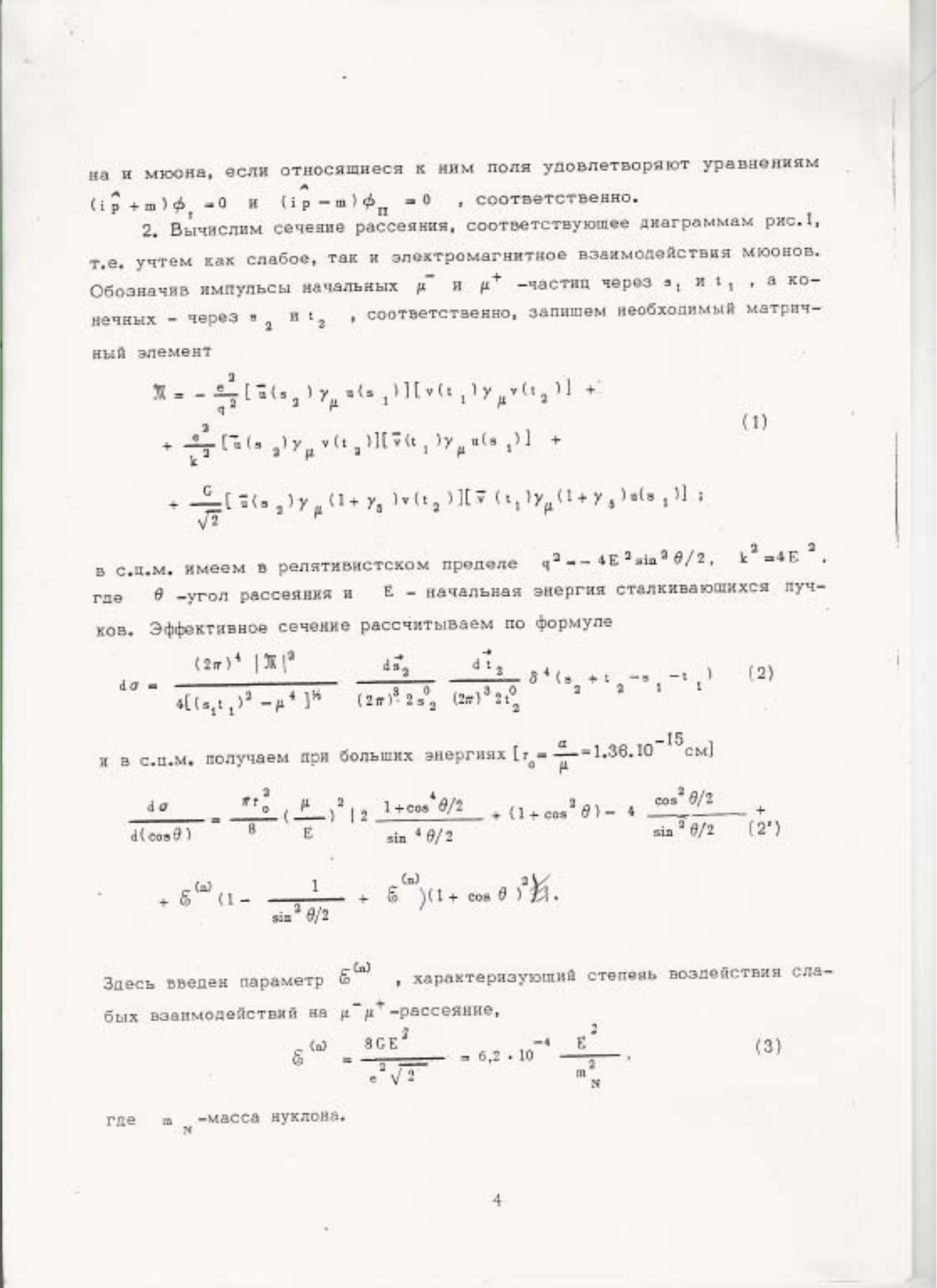}}
\end{center}
\end{figure}

\newpage
\begin{figure}[htb]
\begin{center}
\resizebox{16cm}{!}{\includegraphics*{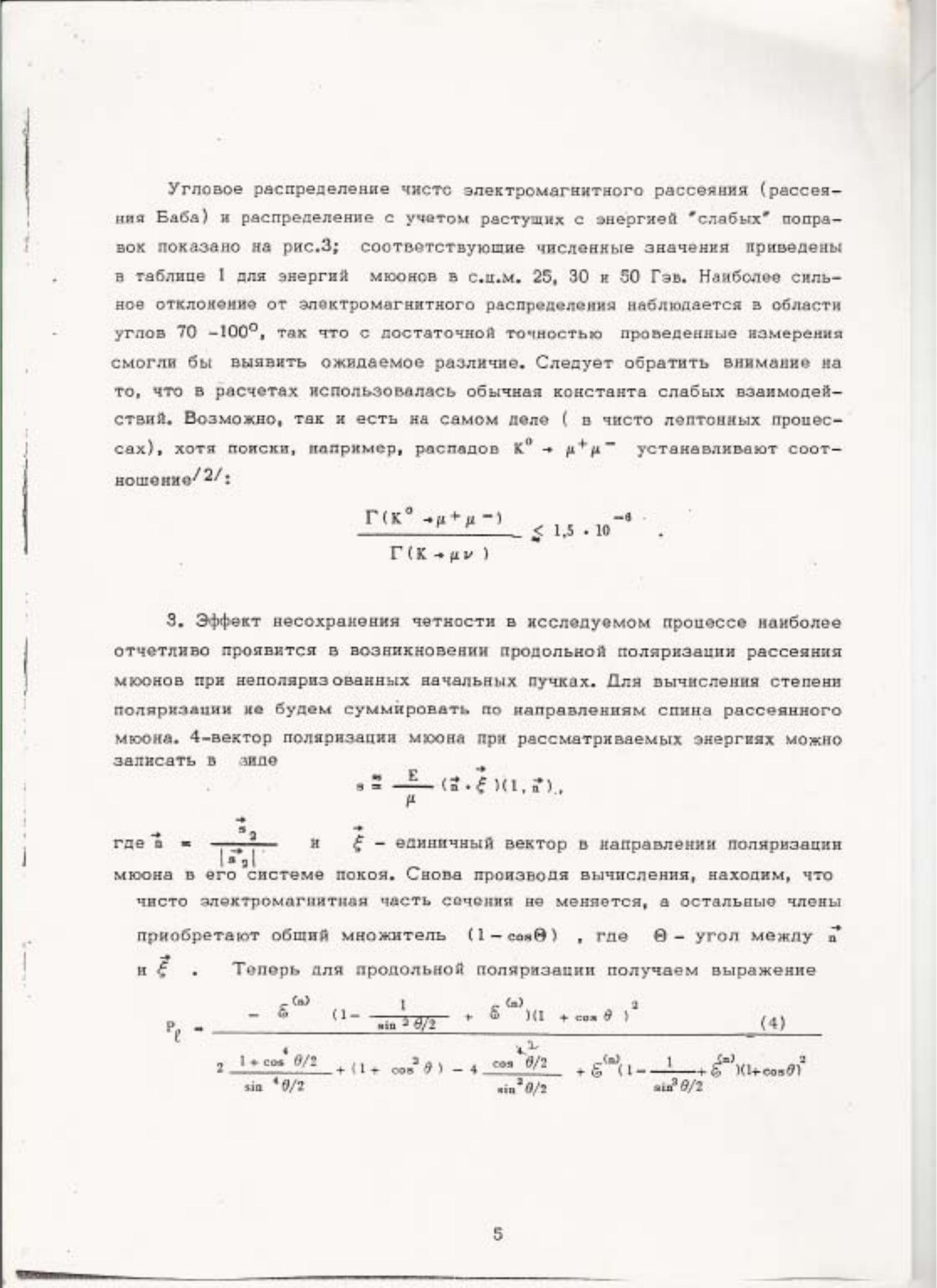}}
\end{center}
\end{figure}

\newpage
\begin{figure}[htb]
\begin{center}
\resizebox{16cm}{!}{\includegraphics*{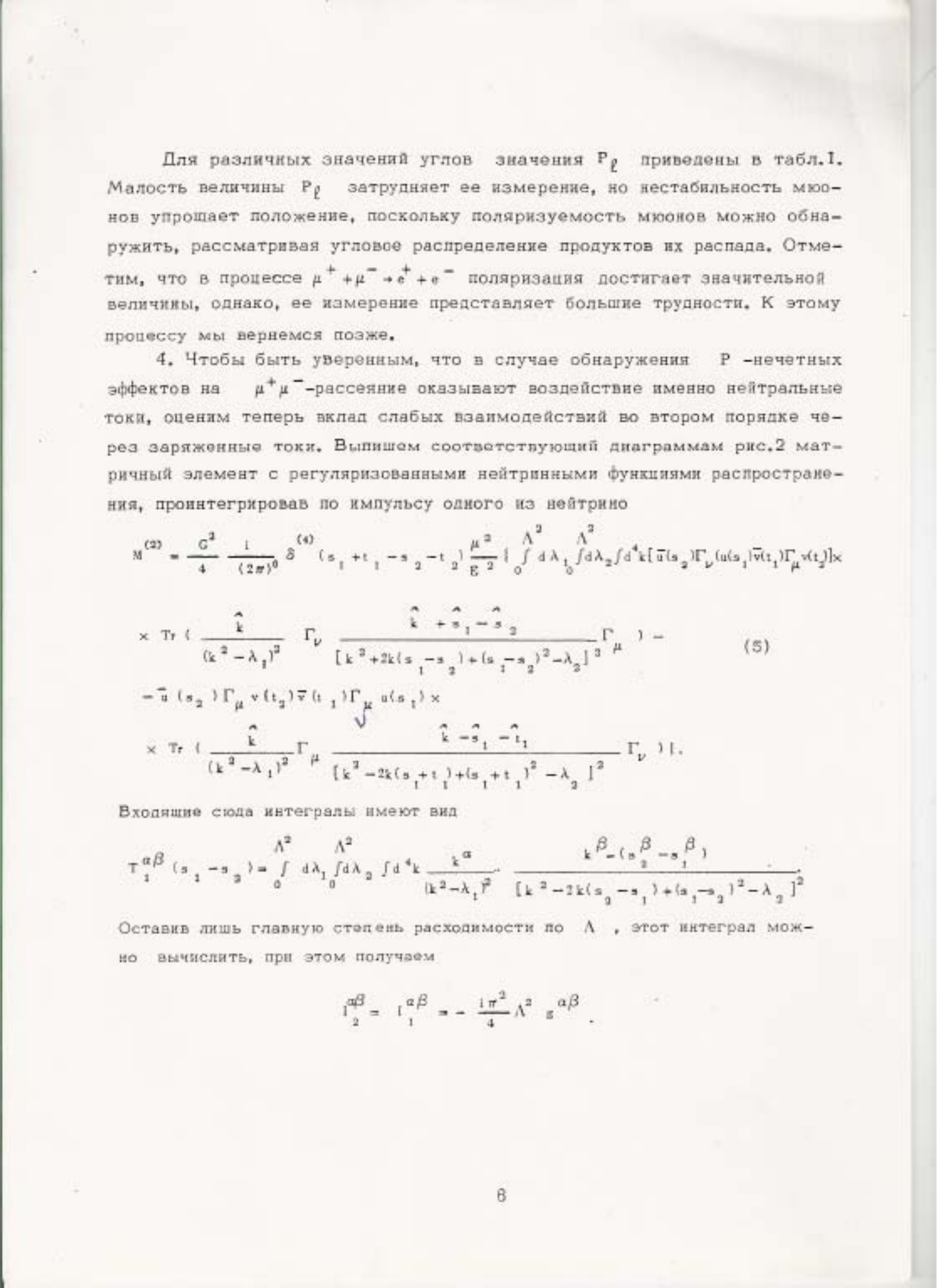}}
\end{center}
\end{figure}

\newpage
\begin{figure}[htb]
\begin{center}
\resizebox{16cm}{!}{\includegraphics*{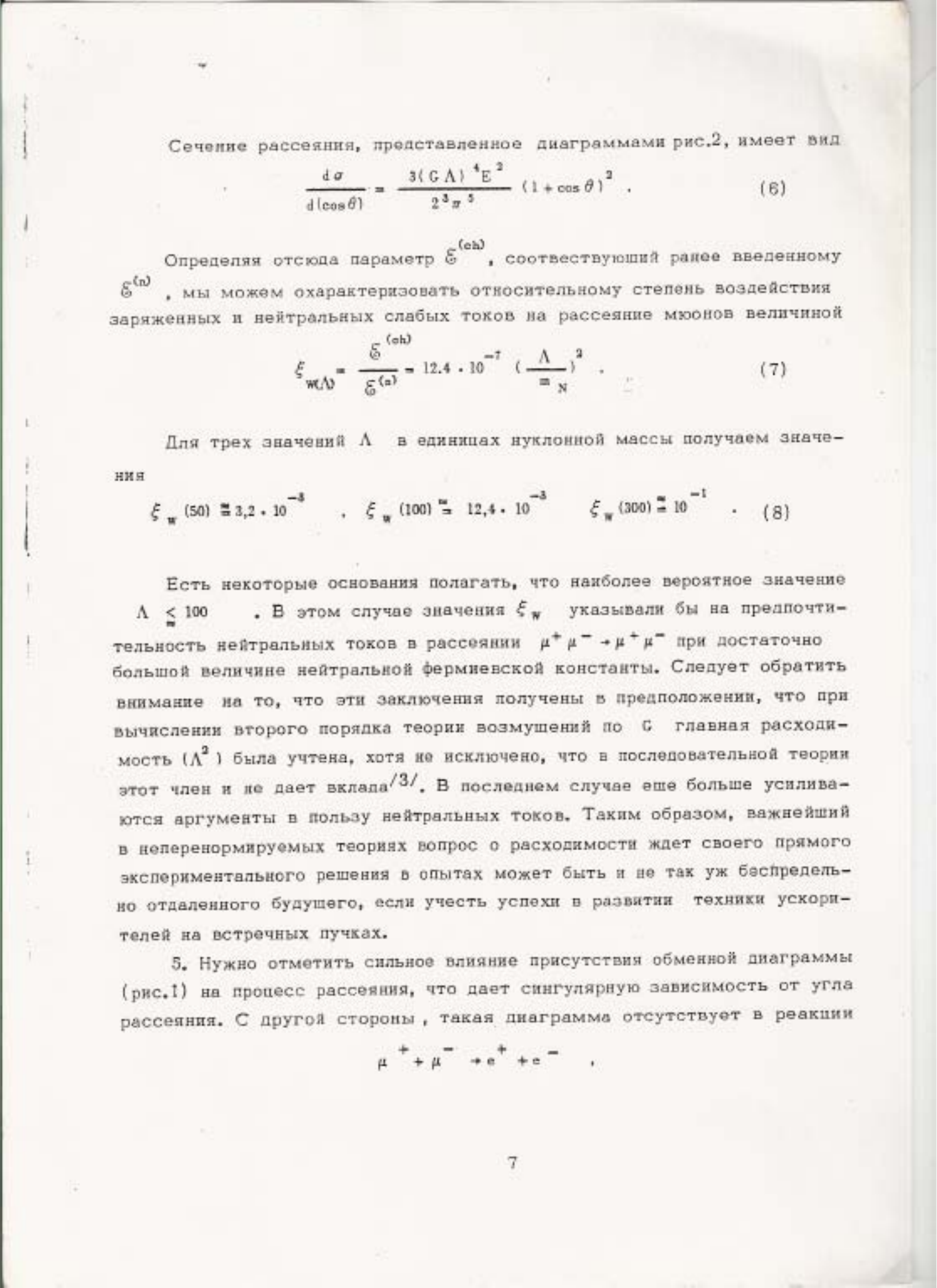}}
\end{center}
\end{figure}

\newpage
\begin{figure}[htb]
\begin{center}
\resizebox{16cm}{!}{\includegraphics*{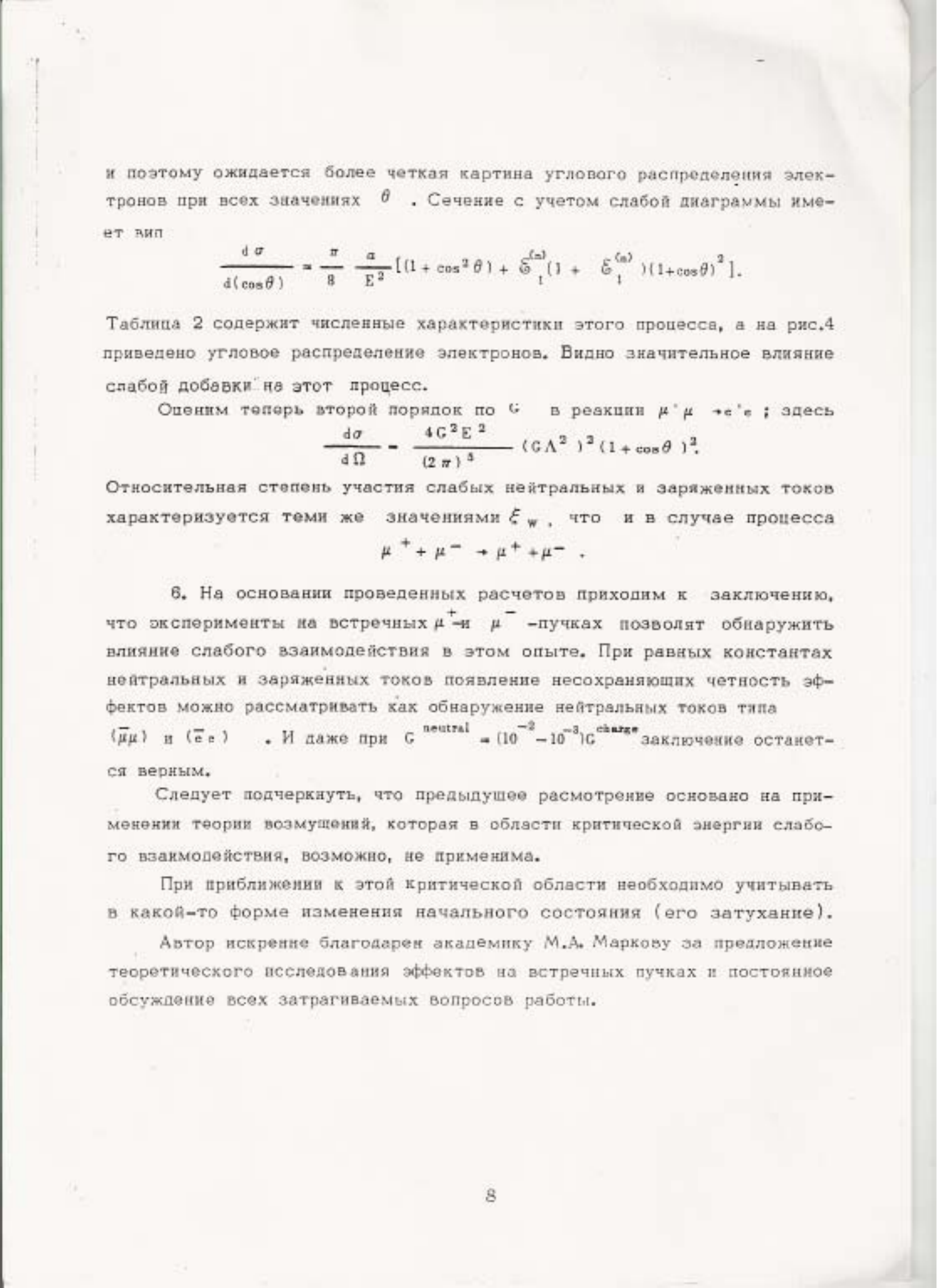}}
\end{center}
\end{figure}

\newpage
\begin{figure}[htb]
\begin{center}
\resizebox{16cm}{!}{\includegraphics*{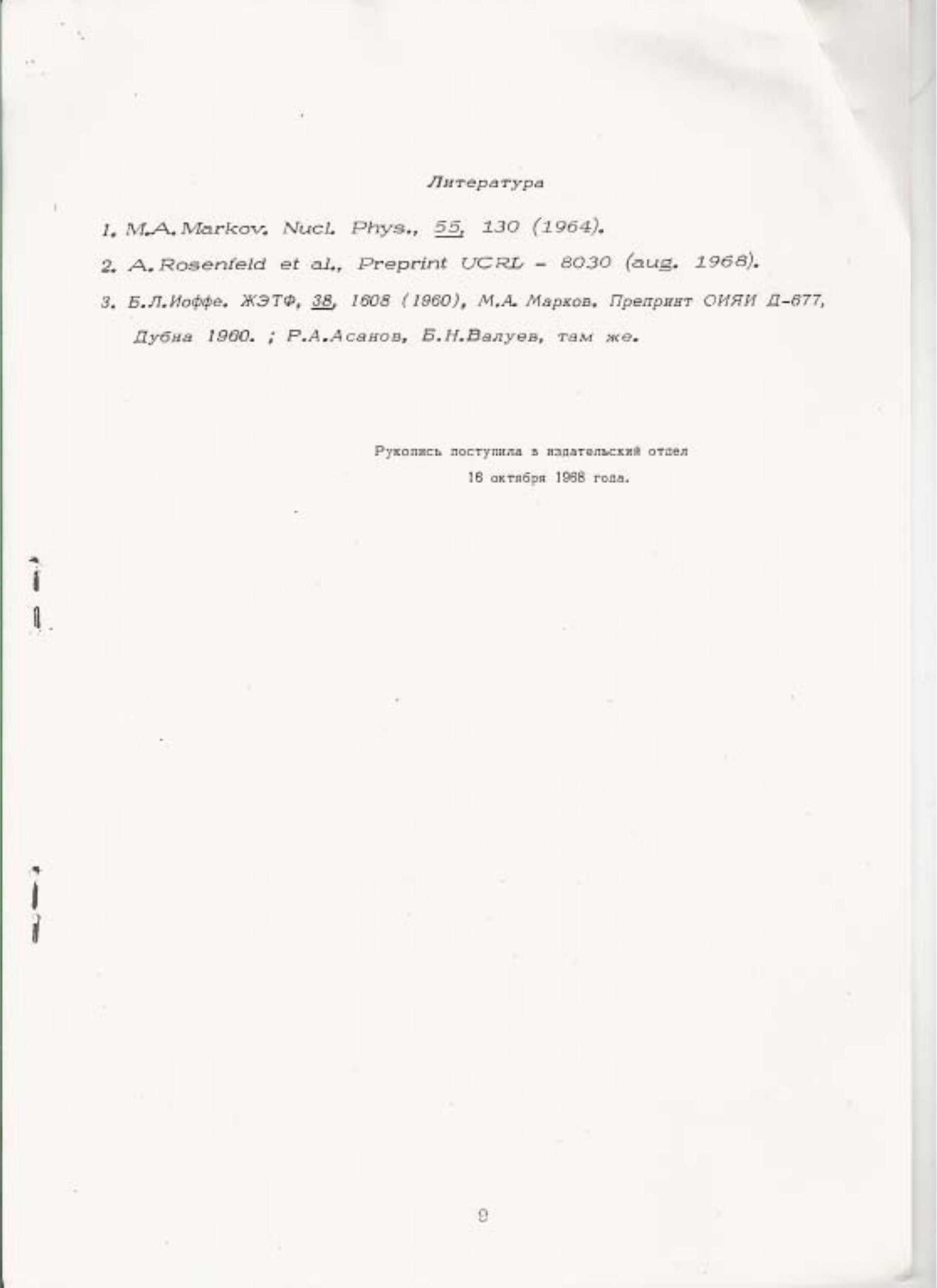}}
\end{center}
\end{figure}


\newpage
\begin{figure}[htb]
\begin{center}
\resizebox{16cm}{!}{\includegraphics*{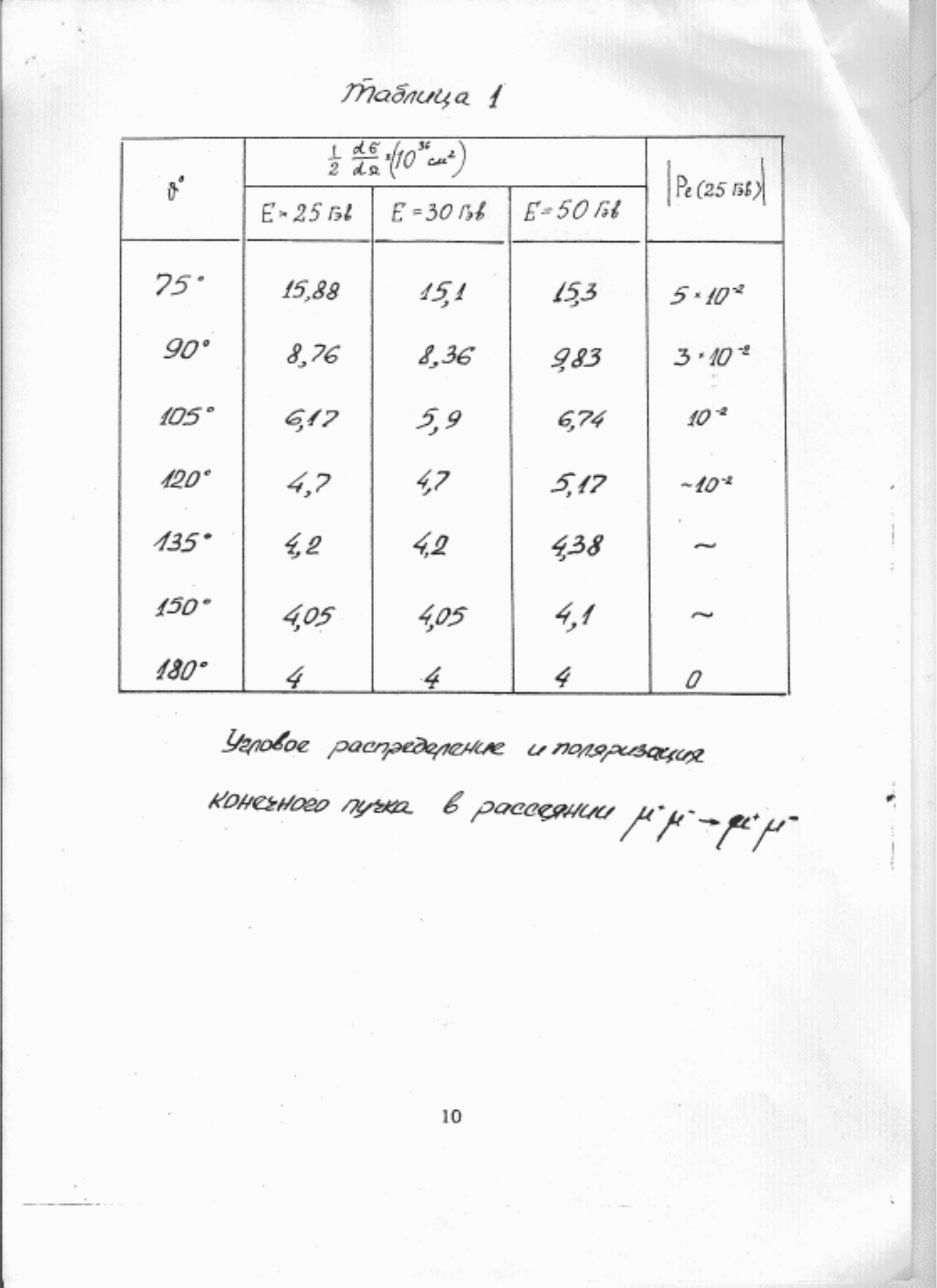}}
\end{center}
\end{figure}

\newpage
\begin{figure}[htb]
\begin{center}
\resizebox{16cm}{!}{\includegraphics*{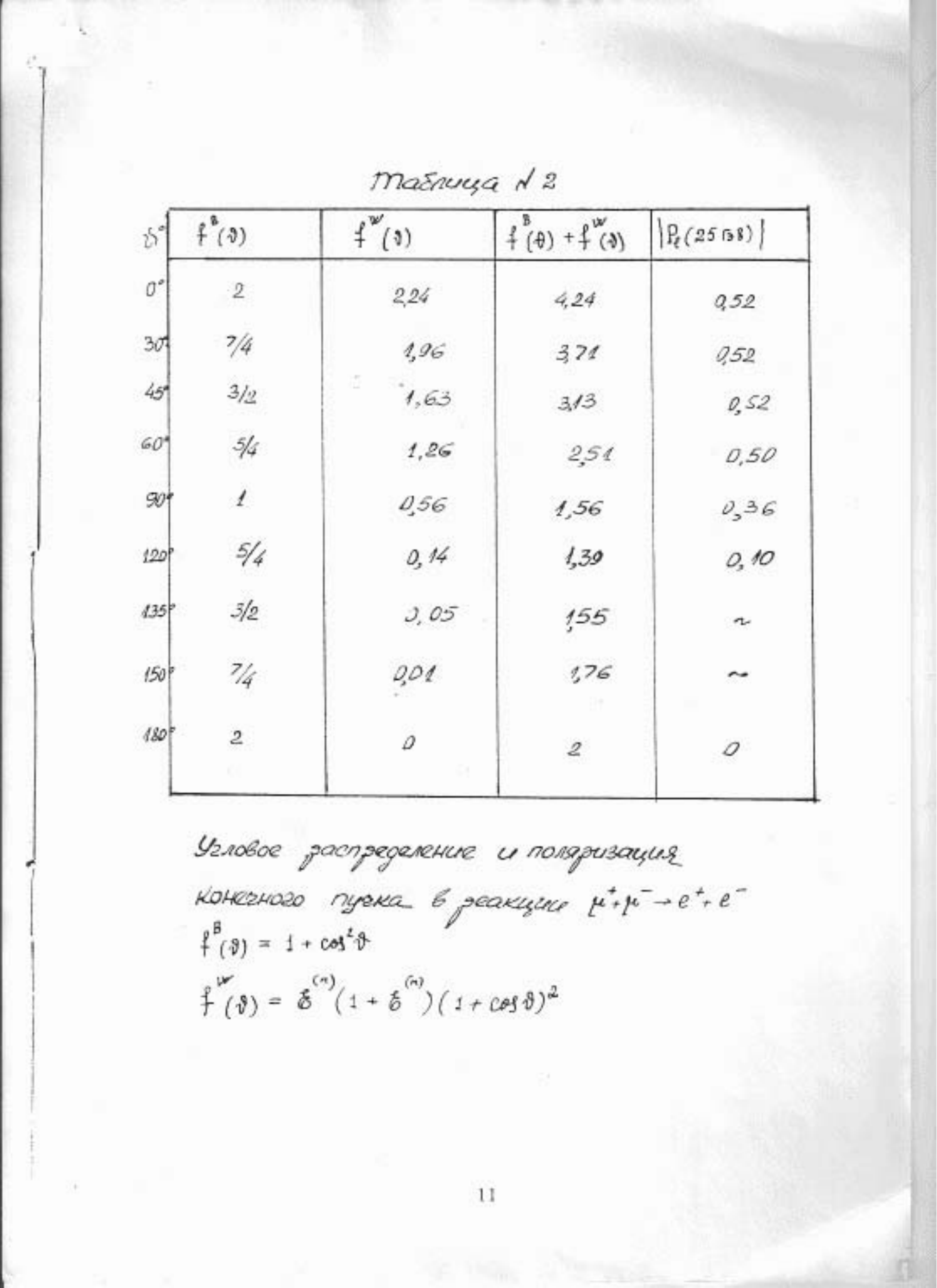}}
\end{center}
\end{figure}


\newpage
\begin{figure}[htb]
\begin{center}
\resizebox{16cm}{!}{\includegraphics*{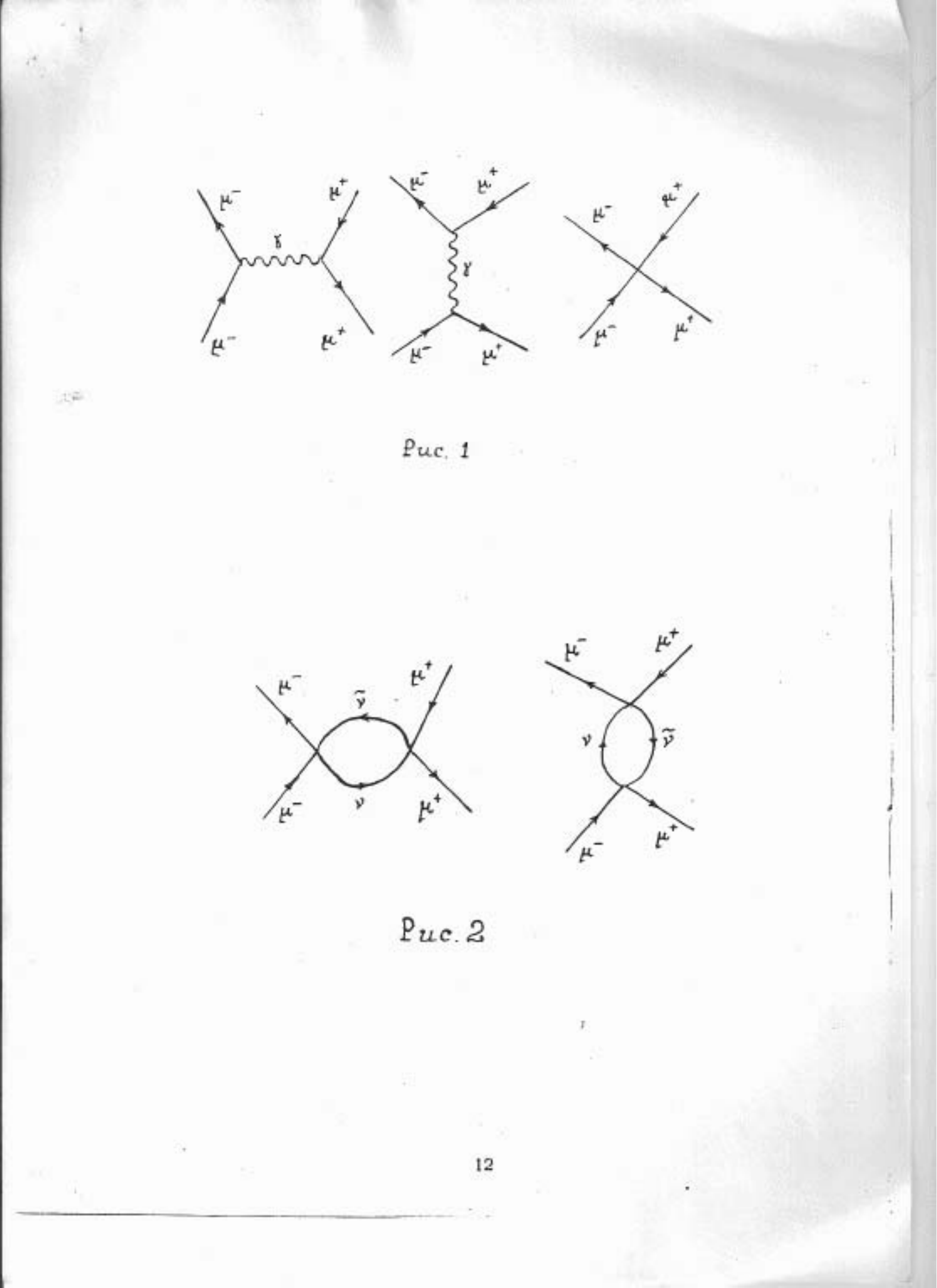}}
\end{center}
\end{figure}

\newpage
\begin{figure}[htb]
\begin{center}
\resizebox{16cm}{!}{\includegraphics*{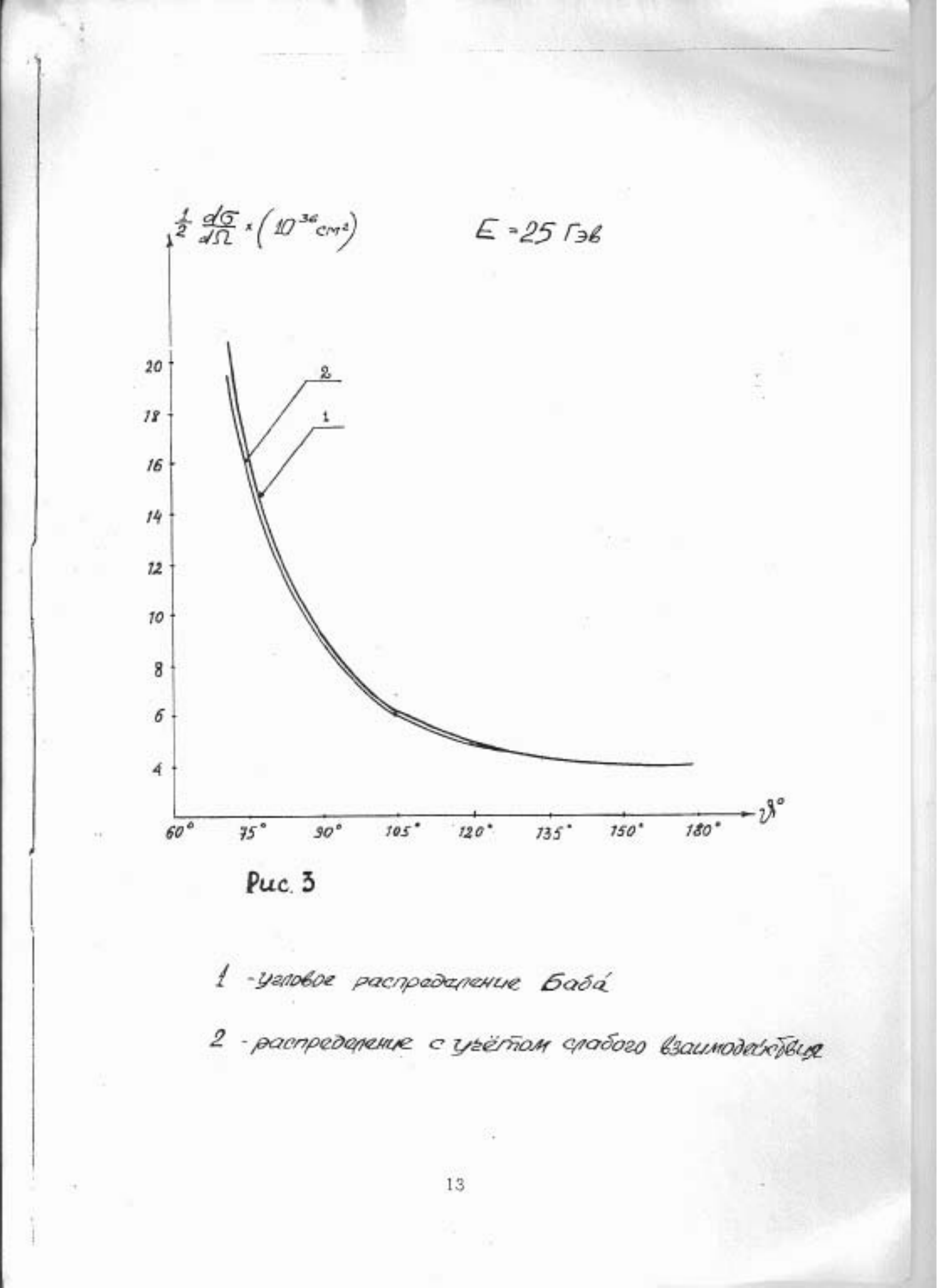}}
\end{center}
\end{figure}

\newpage
\begin{figure}[htb]
\begin{center}
\resizebox{16cm}{!}{\includegraphics*{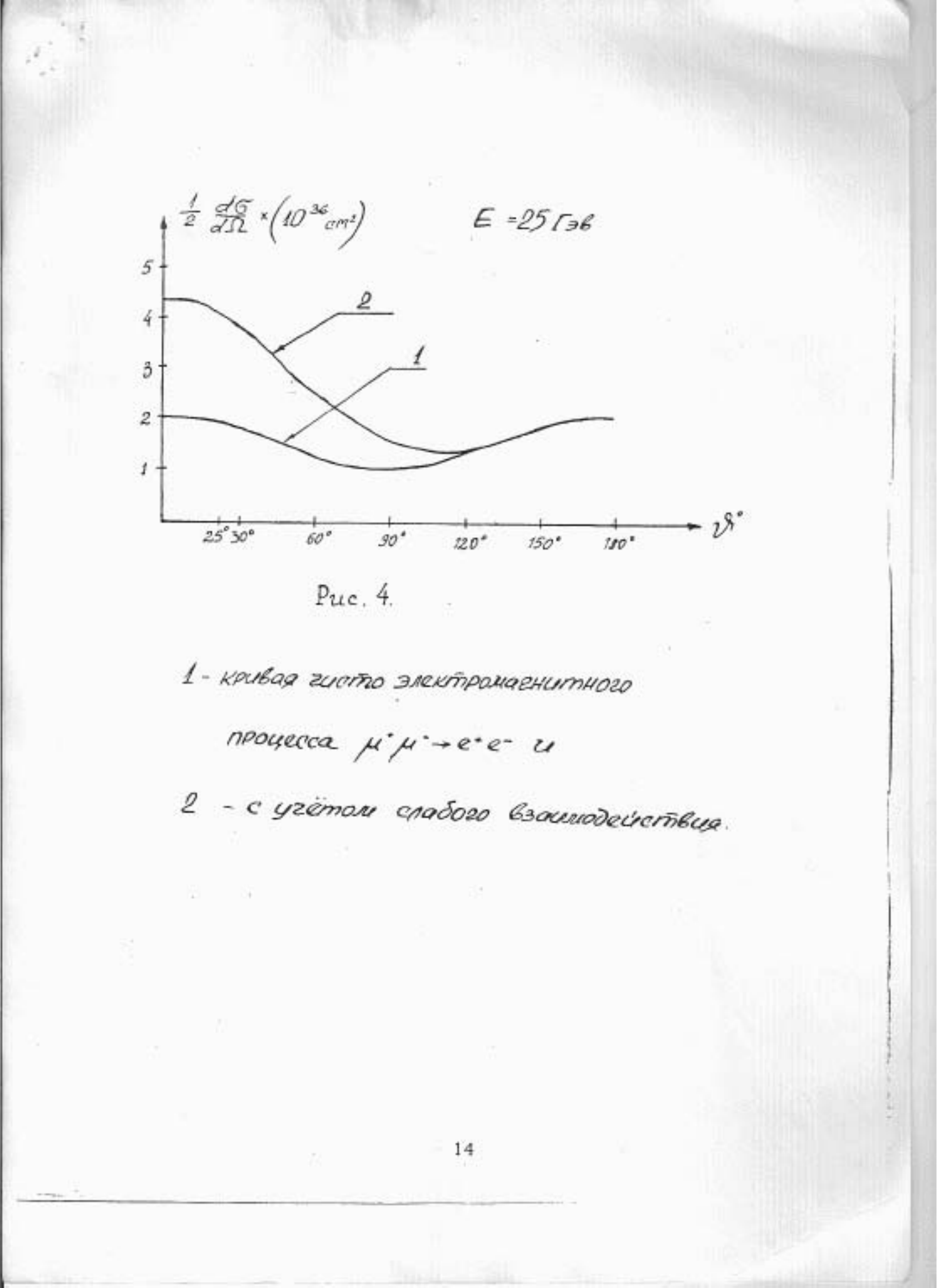}}
\end{center}
\end{figure}

\end{document}